\documentclass[twocolumn,twoside]{revtex4}
\usepackage{graphicx}
\usepackage{amsmath}
\usepackage{amssymb}
\usepackage{fancyhdr}
\begin{document}

\title{Future Neutrino Experiments}

%
 physics over the next decade.~

\author{B.~Jamieson\\
on behalf of the Hyper-Kamiokande Collaboration}
\affiliation{University of Winnipeg, Winnipeg, MB, Canada, R3B 2E9}

\begin{abstract}
The discovery of neutrino oscillations and the large mixing parameters in the Pontecorvo Maki Nakagawa Sakata matrix has opened a window to search for CP-violation in neutrinos.  Two long-baseline neutrino experiments, DUNE and Hyper-Kamiokande, are being prepared in the next decade to search for this CP-violation with intense beams of muon and anti-muon neutrinos.  The JUNO experiment in China will study neutrino oscillations at medium baselines with several goals, including determining the mass ordering of the neutrinos by using the interference of the solar and atmospheric neutrinos.  Short-baseline neutrino oscillation experiments have seen anomalies in their data that could either be explained by sterile neutrinos or new nuclear effects.  Several experiments, including the Fermilab short baseline program, Prospect-II and JUNO-TAO are being planned to understand the source of these anomalies.  This paper provides a short review of these exciting experiments that will lead to new discoveries in neutrino physics over the next decade.
\end{abstract}

\maketitle

\thispagestyle{fancy}


\section{Introduction}
The title of this paper is too broad to be covered in a five-page review, and therefore will not be a complete review.  Instead, the paper will look at the main physics sensitivities of four of the many future neutrino experiments.  First, the paper will focus on a review of the physics sensitivities of the two large long-baseline neutrino oscillation experiments Deep Underground Neutrino Experiment (DUNE)~\cite{dune} and Hyper-Kamiokande (Hyper-K)~\cite{hyperk}.  Second, the JUNO~\cite{juno} physics program will be reviewed, and finally, the Fermilab Short Baseline (SBL) program will be discussed ~\cite{fermilabsbl1,fermilabsbl2}.

First, we consider the Pontecorvo Maki Nakagawa Sakata (PMNS) matrix $U$ describing the mixing of neutrinos.  The PMNS matrix can be written in terms of mixing angles $\theta_{12}$, $\theta_{13}$, and $\theta_{23}$ describing the strength of the oscillations, and CP-violating phases ($\delta$, $\alpha_1$, and $\alpha_2$).  Neutrino oscillation experiments are not sensitive to the so-called Majorana phases ($\alpha_1$ and $\alpha_2$) and will be ignored here.  Neutrinos produced in a particular flavor ($\nu_e$, $\nu_{\mu}$, $\nu_{\tau}$) propagate as a combination of mass states ($\nu_1$, $\nu_2$, $\nu_3$), and the PMNS matrix is:

\begin{align*}
\begin{pmatrix}
\nu_e \\
\nu_{\mu}\\
\nu_{\tau} \\
\end{pmatrix} = &
\begin{pmatrix}
1 & 0 & 0 \\
0 & c_{23} & s_{23} \\
0 & -s_{23} & c_{23}
\end{pmatrix}
\begin{pmatrix}
c_{13} & 0 & s_{13} {\rm e}^{-i \delta} \\
0 & 1  & 0 \\
-s_{13} {\rm e}^{i \delta} & 0 & c_{13}
\end{pmatrix} \times \\
 &
\begin{pmatrix}
c_{12} & s_{12}& 0 \\
-s_{12} & c_{12} & 0 \\
0 & -0 & 1
\end{pmatrix}
\begin{pmatrix}
\nu_1 \\
\nu_2\\
\nu_3 \\
\end{pmatrix}
\end{align*}
where $c_{ij} = \cos{\theta_{ij}}$, and $s_{ij} = \sin{\theta_{ij}}$.

There are three main open questions in neutrino oscillation physics.  The first is whether the oscillation of $\nu_{\mu} \rightarrow \nu_e$ is the same as $\bar{\nu}_{\mu} \rightarrow \bar{\nu}_e$ indicating a CP-violating phase $\delta \ne 0$?  The second is whether $\nu_1$ is the lightest neutrino giving us a Normal Hierarchy (NH), or $\nu_3$ is the lightest neutrino giving us the Inverted Hierarchy (IH)?  Finally, there is interest in knowing whether the mixing of the atmospheric neutrinos is maximal giving $\theta_{23} = 45^\circ$ or if it falls below or above this angle ($\theta_{23}$ octant)?  The next generation of long and medium baseline experiments (Hyper-K, DUNE, and JUNO) have a good chance of answering these open questions.

\section{DUNE}

\begin{figure}[h]
\centering 
\includegraphics[clip, trim=1.0cm 1.0cm 9.1cm 0.5cm, width=60mm]{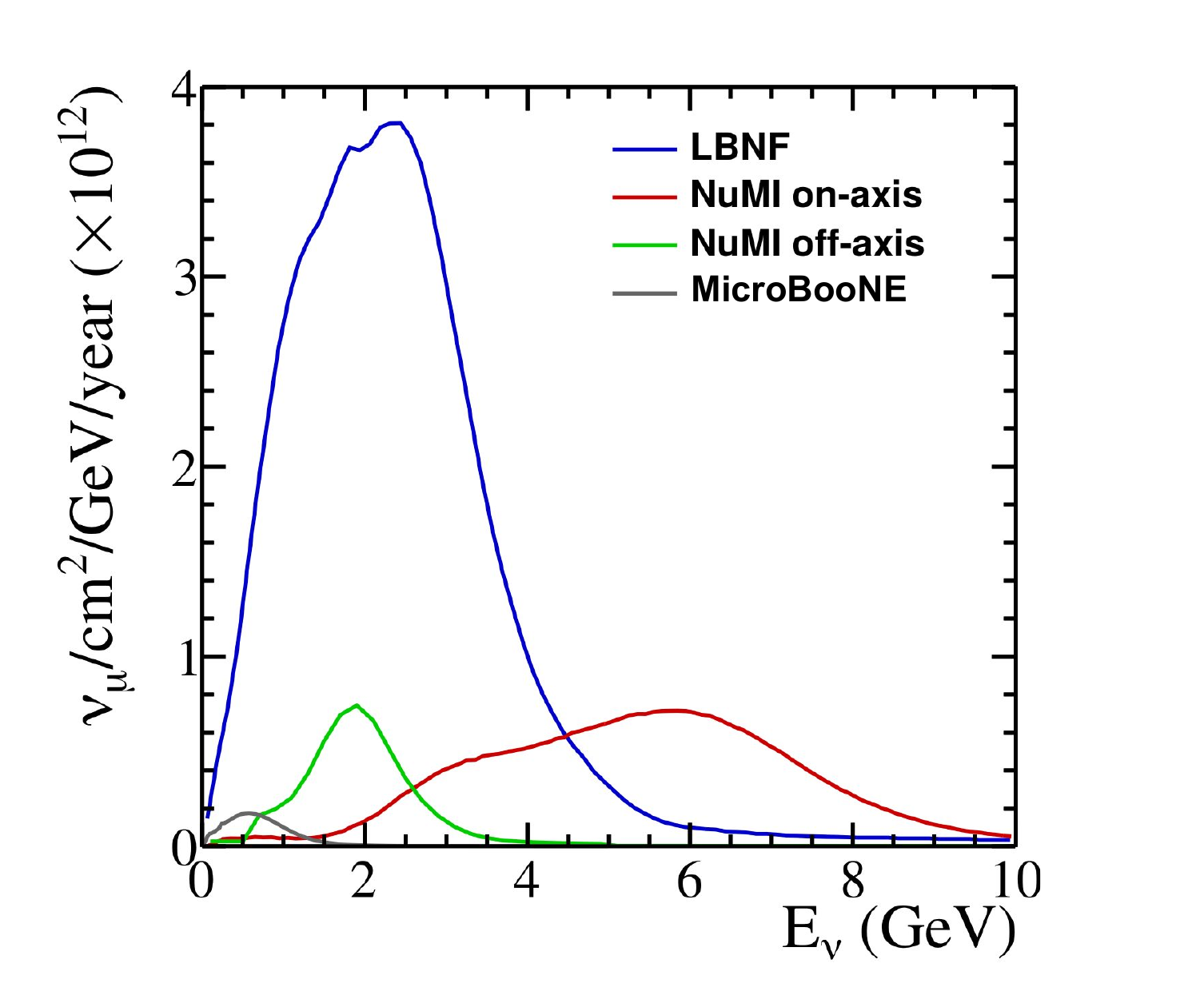}
\caption{DUNE beam flux compared to current generation neutrino beams at Fermilab (Figure from C. Wilkinson, DUNE collaboration).} \label{fig:dunebeam}
\end{figure}

DUNE is a 1300~km baseline neutrino experiment under construction that consists of a 1.2 MW $\nu_{\mu}$ or $\bar{\nu}_{\mu}$ neutrino beam at Fermilab (the PIP-II beamline), a near detector complex, and four 17 kT liquid Ar Time Projection Chambers (LArTPCs) located 1.5~km underground at the 1300~km baseline at Sanford Underground Research Facility. The unoscillated beam flux prediction is shown in FIG.~\ref{fig:dunebeam} along with the flux from other current generation neutrino beams at Fermilab. 

\begin{figure*}[ht]
\centering
\includegraphics[clip, trim=0.1cm 3.1cm 0.1cm 0.1cm, width=160mm]{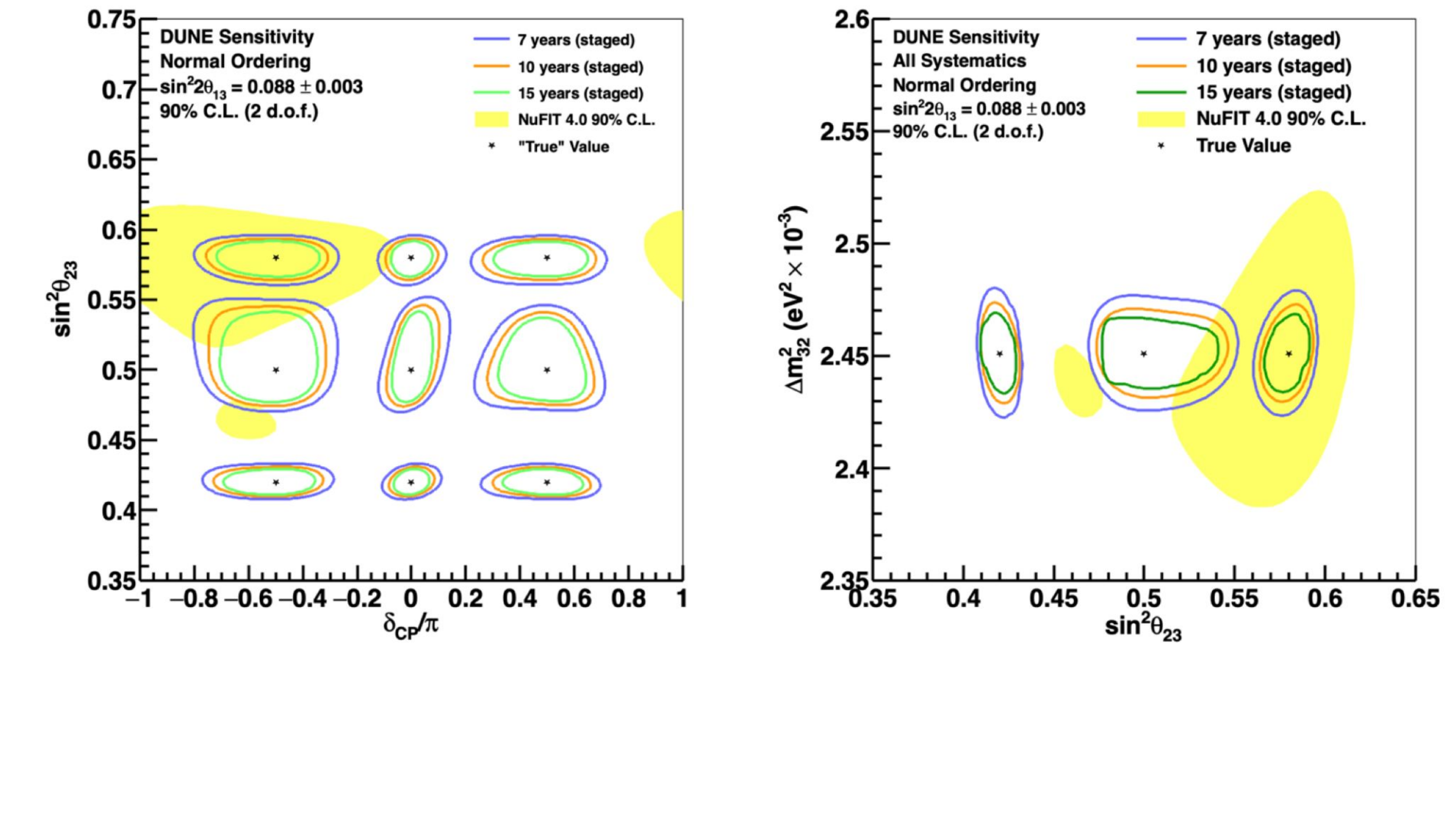}
\caption{DUNE disappearance sensitivity (left), and appearance sensitivity (right) for different potential values of the neutrino oscillation parameters. Figures prepared by the DUNE collaboration~\cite{pickering}.}\label{fig:dunesensitivities}
\end{figure*}

The DUNE near detector complex will consist of a LArTPC, muon-spectrometer and a beam monitor.  The LArTPC and muon-spectrometer will be movable along 28.5 m long rails, allowing the neutrino spectrum to be measured for off-axis angles between zero and $\sim 3.3^\circ$ with peak neutrino energies from 0.5~GeV to 2.0 GeV. Linear combinations of these neutrino spectra can be used to obtain the lepton kinematics of the oscillated flux times cross-section at the far detector. By using the same nuclear target (Ar) as the far detector, and using this prism technique, the systematic uncertainties in neutrino oscillation parameter determination can be reduced. 

Recent talks from DUNE collaborators show plots of the appearance sensitivity and disappearance sensitivity for different numbers of years running.  The appearance and disappearance sensitivities for DUNE are shown in FIG.~\ref{fig:dunesensitivities}.

\section{Hyper-Kamiokande}

\begin{figure}[h]
\centering 
\includegraphics[width=70mm]{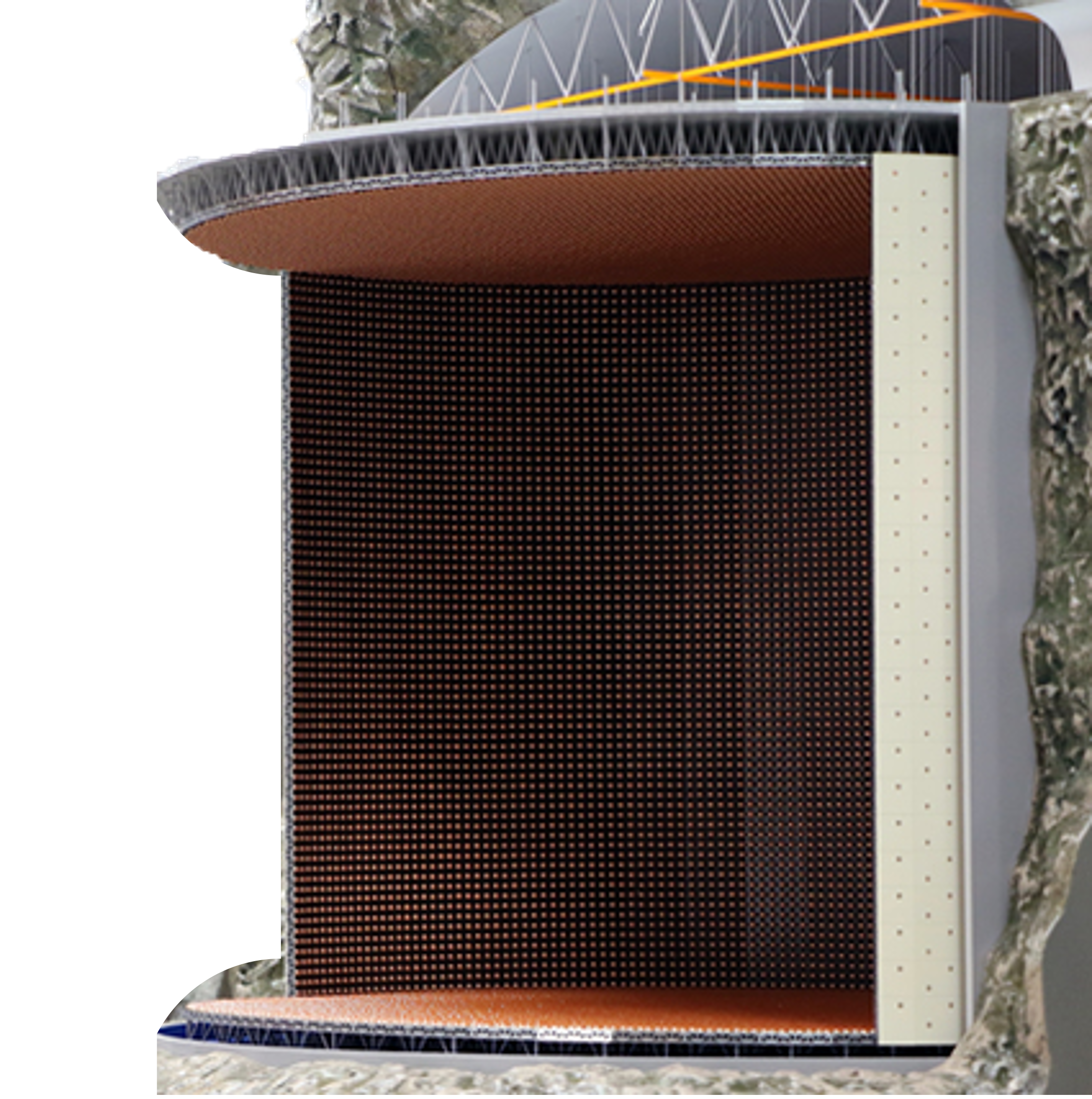}
\caption{Conceptual drawing of the Hyper-Kamiokande water Cherenkov far detector (Figure from Hyper-K Collaboration).} \label{fig:hyperk}
\end{figure}

The Hyper-K experiment is a world-leading neutrino experiment, building on the success of Super-Kamiokande and T2K.  The experiment has a broad and ambitious physics program covering many neutrino sources, as well as proton decay measurements.  The experiment will consist of upgraded 1.2~MW $\nu_{\mu}$ and $\bar{\nu}_{\mu}$ beam at the Japan Proton Accelerator Research Complex (J-PARC) on the East coast of Japan, a near detector complex including a new Intermediate Water Cherenkov Detector (IWCD), and the Hyper-Kamiokande water Cherenkov far detector, shown in Fig.~\ref{fig:hyperk}. A key improvement to Hyper-K over Super-K is the improved PMTs with nearly doubled quantum efficiency and the reduction in the timing resolution (making the timing resolution twice as good).

The Hyper-K detector will have an 8$\times$ larger fiducial mass over Super-Kamiokande.  Hyper-K will: be 71 m tall and 68 m in diameter; have a 1 m wide veto region on the sides; and have a 2 m veto region on the top and bottom.  The 20,000 box-and-line 50 cm PMTs will provide 20\% photo coverage with 1.5 ns timing resolution and double the quantum efficiency of the Super-K PMTs.  Additional photo coverage will come from multi-PMT modules consisting of 19 eight-inch PMTs.  The mPMT modules will provide improved position, timing, and direction resolution to act as in-situ calibration of the 50 cm PMTs.

\begin{figure}[h]
\centering 
\includegraphics[clip, trim=0.3cm 0.3cm 0.3cm 0.2cm, width=80mm]{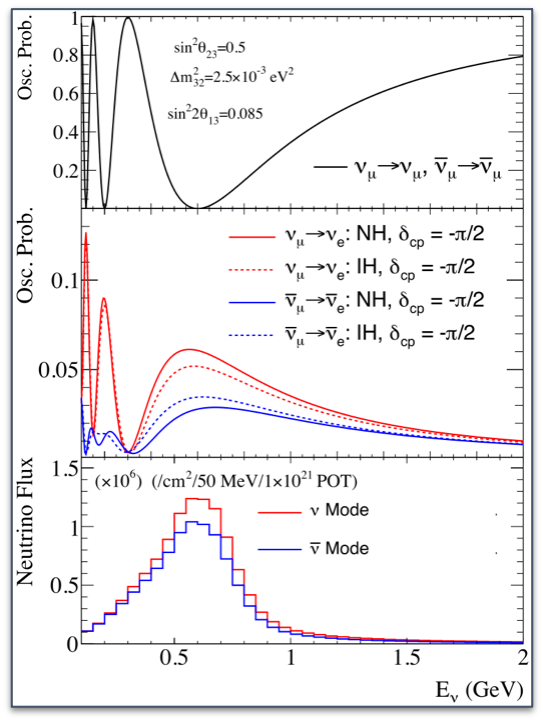}
\caption{Hyper-K beam flux (bottom) compared to the oscillation probabilities for $\nu_{\mu}$ disappearance (top), $\nu_e$ appearance (middle) (Figure from Hyper-K collaboration).} \label{fig:hkbeam}
\end{figure}

The Hyper-K beam flux and oscillation probabilities as a function of neutrino energy are shown in FIG.~\ref{fig:hkbeam}. By using a $2.5^{\circ}$ off-axis beam, the beam flux peaks at the oscillation maximum, and it subtends a smaller range of energies.

Upgrades to the T2K off-axis near detector which includes a new Super Fine Grained Detector (SFGD) with over two million scintillator cubes between two horizontally stacked TPC modules to allow measurements of neutrino interaction products at larger angles from the beam direction in a magnetized detector.  

A new Intermediate Water Cherenkov Detector (IWCD) at $\sim1$~km from the beam target will be built, which uses the same nuclear target as the far detector, and uses the prism technique to measure different beam fluxes for beams between $1^{\circ}$ and $3^{\circ}$ by raising and lowering an instrumented region in a pit of water.  IWCD will use finer-grained photosensors in multi-PMT modules to improve timing and fiducial volume resolution.  By using linear combinations of the measurements at different off-axis angles, either a monochromatic beam or a beam matching the oscillated flux can be mimicked in the oscillation analysis to reduce the flux $\times$ cross-section uncertainties by directly using the measured lepton kinematics.

\begin{figure}[htb]
\centering 
\includegraphics[width=80mm]{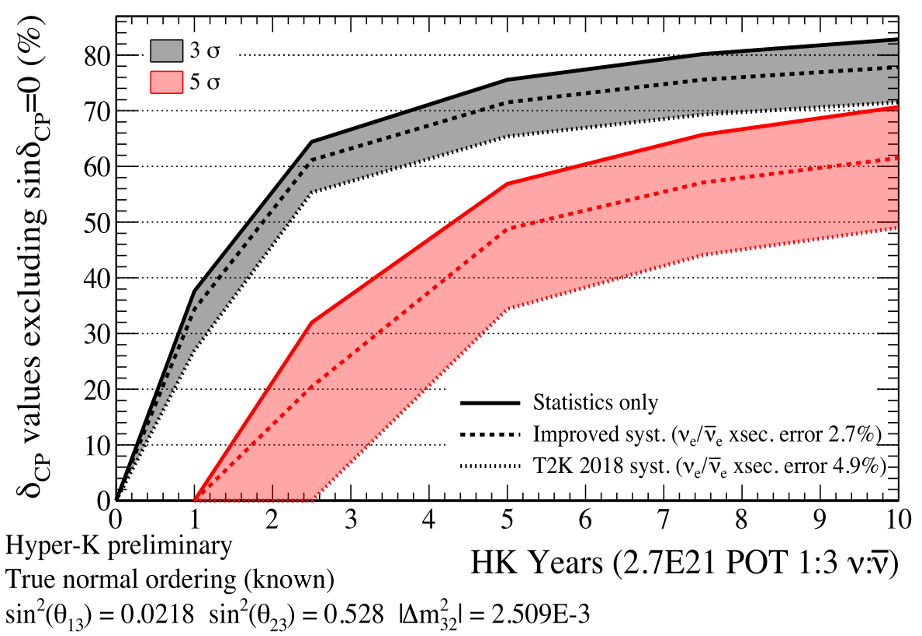}
\caption{Hyper-K percentage of $\delta_{CP}$  values excluding $\sin{\delta_{CP}}=0$, where the shaded bands show the effect of improving systematic uncertainties (Figure from Hyper-K collaboration).} \label{fig:hkdeltacp}
\end{figure}

One of the key goals of Hyper-K is the discovery of a non-zero CP-violating phase in neutrino oscillation.  How quickly this discovery can be made depends on improvements to systematic uncertainties as can be seen in the plot of the fraction of $\delta_{CP}$ values excluding $\sin{\delta_{CP}}=0$ shown in FIG.~\ref{fig:hkdeltacp}.  Without improving over the systematic uncertainties of T2K, there could be the discovery of a non-zero $\delta_{CP}$ within 5 years of running for 50\% of the phase-space of values.

In addition to the CP-violation search, Hyper-K will improve measurements of solar neutrinos, atmospheric neutrinos, and searches for nucleon decay.  Hyper-K will also act as a detector for supernova neutrinos, where of order 100,000 neutrinos from supernova within the distance to the galactic center would be detected, allowing for discrimination between different models of supernovas.  Finally, Hyper-K will observe supernova relic neutrinos within 10 years of starting operations.

Hyper-Kamiokande began construction in 2021 with the access tunnel excavation completed on schedule in June 2022.  The main cavern excavation is underway.  The 50~cm PMT production is on schedule, and inspection and testing are ongoing.  Development of the PMT covers is in progress, as is the development of multi-PMT modules with LED sources of different wavelengths to do detailed detector calibration.  Hyper-K is currently projected to begin operations in 2027.

\section{JUNO}

Jiangmen Underground Neutrino Observatory (JUNO) is a multi-purpose liquid scintillator experiment in China that will be 700 m underground near Jiagnmen city.  JUNO is a 20 kton LAB-based liquid scintillator-based detector with 78\% PMT coverage allowing it to reach a better than $3\%/\sqrt{E({\rm MeV})}$ energy resolution, and calibrations to allow an energy scale uncertainty of less than one percent.  The reactor $\bar{\nu}_e$ will come from two reactor complexes at a baseline of about 52.5 km totaling $\sim 27$ GW$_{th}$ of thermal power.  The detector will detect $\sim$60 reactor anti-neutrinos per day.

\begin{figure}[htb]
\centering 
\includegraphics[width=80mm]{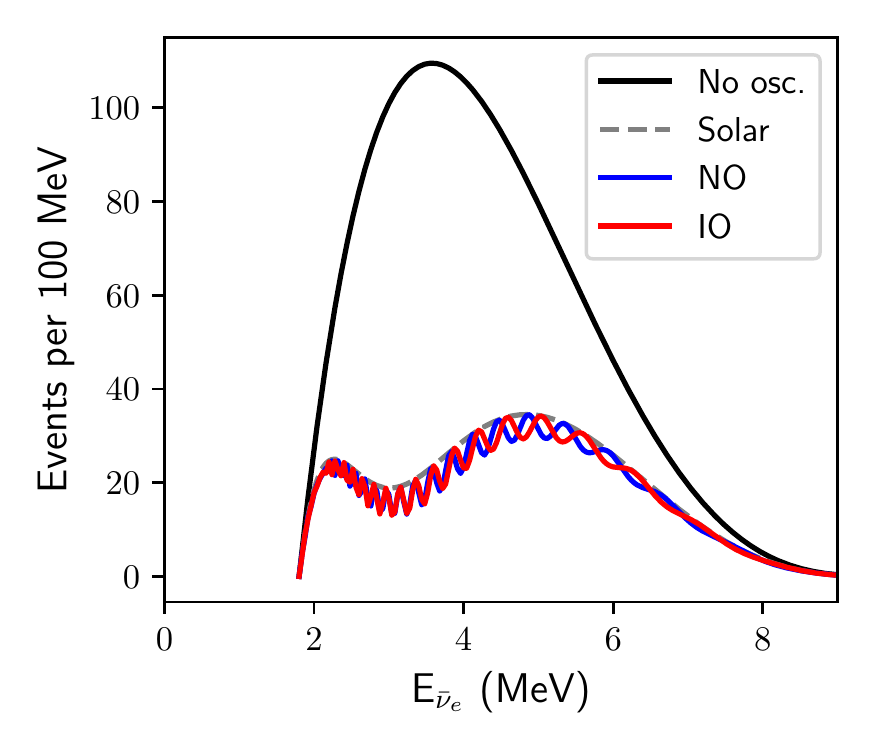}
\caption{JUNO events per 2 days of data taking, showing no oscillation flux (black) the effect of solar oscillations parameters (grey-dashed), and the atmospheric oscillation parameters in the normal ordering (blue) and inverted ordering (red).} \label{fig:junoev}
\end{figure}

For the JUNO baseline, there is an interesting separation of the Normal Ordering (NO) and Inverted Ordering as a phase shift in the oscillations as seen in FIG.~\ref{fig:junoev}.  The $\bar{\nu}_e$ disappearance oscillation probability is approximated as 
\begin{align*} 
P_{\bar{\nu}_e \rightarrow \bar{\nu}_e} = & 1 - \cos^4{2 \theta_{13}} \sin^2{2\theta_{12}} \sin^2{\frac{\Delta m^2_{12} L}{4E}} \\ 
  & - \sin^2{2\theta_{13}} \left( \cos^2{\theta_{12}} \sin^2{ \frac{\Delta m^2_{31} L}{4E} } \right. \\
  & \left. + \sin^2{\theta_{12}} \sin^2{\frac{\Delta m^2_{32} L}{4E} }   \right) . 
 \end{align*}
The first two terms give the oscillation from the solar only parameters, while the last term adds in the effect from the atmospheric oscillations.
 
JUNO will be able to measure the solar ($\sin^2{\theta_{12}}$, $\Delta m^2_{21}$) and atmospheric ($|\Delta m^2_{32}|$) neutrino oscillation parameters with a precision of 0.6\% or better with six years of operation.  JUNO is likely to determine the neutrino mass ordering as it was designed to do before the long-baseline experiments begin their measurements.

\section{Fermilab short baseline neutrinos}

At short baselines, several neutrino experiments are showing unexpected results at $L/E \sim 1$~km/GeV.  The $\bar{\nu}_e$ disappearance has been observed in reactor experiments~\cite{abazajian} with a ratio $R=N_{exp}/N_{calc}=0.934\pm0.024$ at distances of less than 1~km across several reactor experiments.  In $\nu_e$ source calibrations of solar neutrino experiments there is an observed disappearance with $R=N_{exp}/N_{calc}=0.84\pm0.05$~\cite{arxiv1906.01739}.  The most famous examples are the LSND and MiniBoone results which have an appearance of $\nu_e$ and $\bar{\nu}_e$ neutrino beam experiments\cite{miniboone}. The observations could be explained by sterile neutrinos with $\Delta m^2 \sim 1$~eV$^2$, an unaccounted-for nuclear effect, or other interpretations beyond the standard model. The Fermilab short baseline program will probe these questions using a set of three liquid-argon TPCs which have excellent shower reconstruction and electron / gamma-ray separation.  The near Short Baseline Neutrino Detector (SBND) is 110~m from the Booster Neutrino Beam's target, MicroBooNE is 470~m from the target, and ICARUS is 600 m from the target.

With three years of data from these three detectors, the sterile neutrino 3+1 parameter space can be covered at the 3-5~$\sigma$ level.  With the detailed interaction data of the LArTPCs nuclear effects can be studied with greater certainty, and by using the same nuclear target for all three experiments, systematic uncertainties can be kept to a minimum.

\section{Conclusion}

There are many exciting neutrino experiments under construction around the world, many of which could not be covered in this short conference proceedings.  With these experiments, the next decade of measurements could see us determine if there is CP violation in neutrinos, determine the mass hierarchy of neutrinos, determine the octant of $\theta_{23}$, get closer to measuring the absolute neutrino mass, and make progress on understanding short baseline neutrino anomalies.  With the higher precision experiments being prepared we will likely discover new puzzles related to the properties of neutrinos.

\begin{acknowledgments}
This document is adapted from the ``Instruction for producing FPCP2003
proceedings'' by P.~Perret and from eConf templates~\cite{templates-ref}.  I would like to thank Dr. Anne Schukraft for use of her slides on the Fermilab short baseline program, Dr. Luke Pickering for his slides on DUNE, and Dr. Jinnan Zhang for his slides on JUNO.  I would also like to thank my collaborators on Hyper-Kamikande for selecting me to present this talk, and for their tireless efforts in preparing a world class experiment.
\end{acknowledgments}

\bigskip 

\end{document}